\documentstyle[12pt]{article}

\newcommand {\beq}{\begin{equation}}
\newcommand {\eeq}{\end{equation}}
\newcommand {\beqa}{\begin{eqnarray}}
\newcommand {\eeqa}{\end{eqnarray}}
\newcommand {\beqan}{\begin{eqnarray*}}
\newcommand {\eeqan}{\end{eqnarray*}}
\newcommand {\n}{\nonumber \\}

\newcommand {\Romannumeral}[1]{\uppercase\expandafter{\romannumeral#1}}

\newcommand \vev[1]{\hbox{$\left< \,{#1} \, \right>$}}

\begin{document}
\setlength{\oddsidemargin}{0cm}
\setlength{\baselineskip}{7mm}

\begin{titlepage}
 \renewcommand{\thefootnote}{\fnsymbol{footnote}}
    \begin{normalsize}
    \begin{flushright}
                 UT-675\\
                 KEK-TH-393\\
                 KEK Preprint 94-14\\    
		 hep-lat/9405014 \\
                 April 1994
     \end{flushright}
    \end{normalsize}
    \begin{Large}
       \vspace{1cm}
       \begin{center}
         {\bf The Effect of Dynamical Gauge Field on the 
Chiral Fermion on a Boundary} \\
       \end{center}
    \end{Large}

  \vspace{10mm}

\begin{center}
           Hajime A{\sc oki}$^{1)}$\footnote
           {E-mail address : haoki@danjuro.phys.s.u-tokyo.ac.jp},
           Satoshi I{\sc so}$^{1)}$\footnote
           {E-mail address : iso@danjuro.phys.s.u-tokyo.ac.jp},
           Jun N{\sc ishimura}$^{2)}$\footnote
           {E-mail address : nishi@danjuro.phys.s.u-tokyo.ac.jp,~
JSPS Research Fellow.} {\sc and}  
           Masaki O{\sc shikawa}$^{3)}$\footnote
           {E-mail address : oshikawa@danjuro.phys.s.u-tokyo.ac.jp}\\
      \vspace{1cm}
        $^{1)}$ {\it Department of Physics, University of Tokyo,} \\
               {\it Bunkyo-ku, Tokyo 113, Japan}\\
        $^{2)}$ {\it National Laboratory for High Energy Physics (KEK),} \\ 
               {\it Tsukuba, Ibaraki 305, Japan} \\
        $^{3)}$ {\it Department of Applied Physics, University of Tokyo,} \\
              {\it Bunkyo-ku, Tokyo 113, Japan}\\
\vspace{15mm}

\end{center}

\begin{abstract}
\noindent We study the effect of dynamical gauge field on the Kaplan's chiral 
fermion on a boundary in the strong gauge coupling limit  
in the extra dimension.
To all orders of the hopping parameter expansion, we prove  exact parity 
invariance of the fermion propagator on the boundary. 
This means that the chiral property 
of the boundary fermion, which seems to survive even in the presence of 
the gauge field from a perturbative point of view, 
is completely destroyed by the dynamics of the gauge field. 
\end{abstract}

\end{titlepage}
\vfil\eject

\section{Introduction}
\setcounter{equation}{0}
\hspace*{\parindent}

The construction of a chiral gauge theory on a lattice 
is one of the long--lasting problems in field theory. 
The difficulty can be summarized in the famous no--go theorem by Nielsen and 
Ninomiya \cite{NN}, 
which states that on a $2n$--dimensional lattice the theory has to be 
vector--like under some fundamental assumptions such as locality, positivity, 
hermiticity and charge conservation. 

Among various attempts to construct a chiral gauge theory, 
Kaplan's proposal \cite{Kaplan} seems one of the most promising ones at present 
and has been studied intensively for these two years \cite{works}. 
The idea is to consider a $(2n+1)$--dimensional Dirac fermion 
with a  domain--wall type mass
in the extra dimension, which yields a chiral zeromode 
on the domain wall. 
Equivalently one may consider a $(2n+1)$--dimensional Dirac fermion with 
an ordinary constant mass but on the space with a boundary \cite{Shamir}, 
which plays the same 
role as the domain wall in the Kaplan's model. 
Our argument in this letter can be applied equally well to both the models, 
but to make the explanation as simple as possible, we state only for the 
case of the latter model. 

To be concrete, let us consider a $(2n+1)$--dimensional system with a boundary 
and denote the $2n$--dimensional coordinate by $x^{\mu}~(\mu=0,1,...,2n-1)$ 
and the extra dimensional coordinate by $s$ .
When one considers the system to have a finite extension in the extra dimension
$0\leq s \leq L$, one has two chiral zeromodes with 
opposite handedness on the boundaries $s=0$ and $s=L$.
This fact can be regarded as a natural conclusion of Nielsen--
Ninomiya's theorem. 
But since the chiral fermions with opposite handedness 
live apart in the extra dimension, they are decoupled 
as long as the gauge field is treated as a background.
%
A promising feature of the model is that
the system allows a natural realization 
of fermion number anomaly through the flow of the fermion number current off 
the boundary \cite{Jansen}, in contrast to other attempts such as the one 
using the Wilson--Yukawa system.

The problem is of course what happens if one couples the dynamical $(2n+1)$--dimensional
gauge field to the system.
Let us denote the lattice gauge coupling in the extra dimension $g_s$. 
When $g_s=0$, adopting the axial gauge in the axis of the $s$--direction, 
one has copies of gauge configurations for each  constant--$s$ plane and 
 can prove that the chiral zeromode exists on the boundary.
Moreover the freedom of the gauge field in the extra dimension is completely fixed and the
$(2n+1)$-dimensional system of the gauge field is reduced to $2n$-dimensional, which 
sounds desirable for obtaining a $2n$-dimensional chiral gauge theory.
There seems to be no problem if there is only one boundary at $s=0$, that is,  
if the system is defined from the beginning to extend infinitely 
in the positive $s$--direction \cite{Narayanan}. 
However, if one tries to define the system by first considering a finite 
system 
with the extension $0\leq s \leq L$ and then take the $L\rightarrow \infty$ 
limit, one has another chiral zeromode with the opposite handedness 
on the other boundary at $s=L$, which can never be decoupled 
even in the $L\rightarrow\infty$ limit as long as $g_s=0$. 
The system constructed here is nothing but a $2n$--dimensional vector gauge 
theory. 
Instead of setting $g_s=0$, one might introduce 
some $L$--dependent parameters in the action and 
fine--tune them when one increases $L$ to infinity so that the opposite 
fermions can decouple and the relevant gauge field configurations for the chiral fermion
may become effectively $2n$-dimensional. 
Although this is an interesting possibility,
since we do not know a priori how many
parameters we have to fine--tune, it might require many trials 
and errors in order to see if the scenario really works.

One may take another extreme case,  the large $g_s$ case, where 
the correlation length of the gauge field is finite in the $s$--direction 
and therefore the dynamics of the gauge field can be regarded as 
$2n$--dimensional. 
This limit is against the continuum limit in the $s$ direction but it doesn't lead us
to throw  away this possibility because, in the absence of the gauge field, the existence 
of the chiral zeromode is guaranteed even when the fermion has only finite correlation in the
$s$ direction.
Therefore, if the chiral zeromode survives after taking account of the dynamical gauge field, 
the whole system can be 
regarded as a desirable $2n$--dimensional chiral  gauge theory. 
This possibility has been studied through the mean field analysis \cite{Altes},
but the conclusion is unfortunately a negative one which predicts that 
all the constant--$s$ planes become
independent and that the fermion becomes vector--like in each plane.
>From a perturbative point of view, however,
the presence of the chiral zeromode at the boundary is required from
the induced Chern--Simons action in the bulk \cite{CallanHarvey}, which is
non--renormalizable by the radiative corrections of the gauge field
~\cite{CSnonren}.
Thus the chiral zeromode seems to 
survive even when one considers the dynamics of the gauge field. 

In this letter we study the above problem using the hopping parameter expansion 
in the strong coupling limit. 
Instead of calculating the fermion propagator explicitly, we define parity 
transformation in $2n$--dimensions and see if the fermion propagator is 
invariant under the parity transformation. 
If the fermion propagator is parity invariant, it means that the fermions 
are vector--like, actually much more than that, because it means the 
complete symmetry between the left--handed fermion 
and the right--handed fermion. 
Indeed, when $g_s=\infty$ and neglecting the dynamical fermions, 
we prove that the fermion propagator in each layer is parity invariant.

The paper is organized as follows. 
In Section 2, we define parity transformation in $2n$--dimensions. 
In Section 3, we prove that to all orders of the hopping parameter expansion, 
the fermion propagator is parity invariant. 
Section 4 is devoted to the summary and the discussion. 
\vspace{1cm}

\section{Parity transformation in $2n$-dimensions}
\setcounter{equation}{0}
\hspace*{\parindent}
\label{sec:parity}
In order to obtain a chiral spectrum in $2n$-dimensions, it is necessary
to violate the parity invariance.
The parity transformation in $2n$-dimensions is defined as follows.
\begin{eqnarray}
  \Psi & \rightarrow & \gamma^0 \Psi \nonumber \\
  \bar{\Psi} & \rightarrow & \bar{\Psi} \gamma^0 \\
  x^i  & \rightarrow & - x^i   \;\;\;\;
  \mbox{($i=1,2, \cdots,2n-1$)}, \nonumber
\label{eq:parity}
\end{eqnarray}
where $\Psi$ is the Dirac spinor in $2n$-dimensions.

We can see, for example, that the Dirac Lagrangian in $2n$-dimensions
\begin{equation}
  {\cal L} = i \bar{\Psi} \partial_\mu \gamma^\mu \Psi - m \bar{\Psi} \Psi
\end{equation}
is invariant under the parity transformation.

In the present formulation, the chiral fermion in $2n$-dimensions is
realized on a boundary of $(2n+1)$-dimensional space.
We, therefore, study the parity invariance (in the $2n$--dimensional sense)
of the correlation functions
\begin{equation}
  \vev{ \Psi(x^0,x^i,s=0) \bar{\Psi}(y^0,y^i,s=0)} =
  \vev{ \gamma^0 \Psi(x^0, - x^i,s=0)
                    \bar\Psi(y^0,- y^i,s=0) \gamma^0 }.
\label{eq:corrparity}
\end{equation}
If this condition is satisfied, the spectrum is
completely parity invariant which means that
a chiral fermion cannot appear on the boundary.
\par   
It is instructive to see whether the condition can be satisfied due to
 the symmetry of the action.
When there is no boundary ({\em i.e.} a system of the standard
Dirac fermion in $(2n+1)$-dimensions),
the action has an invariance under the rotation about
the $x^0$-axis by angle $\pi$
\begin{eqnarray}
  \Psi & \rightarrow & \gamma^0 \Psi \nonumber \\
  \bar{\Psi} & \rightarrow & \bar{\Psi} \gamma^0 \\
  x^i  & \rightarrow & - x^i   \;\; \mbox{($i=1,2, \cdots, 2n-1$)} \nonumber \\
  s & \rightarrow & -s. \nonumber \\
\end{eqnarray}
This is nothing but the parity transformation in $2n$-dimension~(\ref{eq:parity}) 
 plus the inversion of the $s$-axis
($s \rightarrow - s $).
Since the inversion of the $s$-axis does not affect the boundary
($s=0$), the condition~(\ref{eq:corrparity}) is satisfied exactly
and there is no chiral mode, as expected.

On the other hand, if there is a boundary,
the action is not invariant under the inversion of the $s$-axis
and we cannot expect that
 the condition~(\ref{eq:corrparity}) is satisfied, so that 
a chiral zeromode can appear on the boundary.
In fact, the presence of the chiral zeromode
can be  explicitly shown when there is no gauge field.
\vspace{1cm}

\section{Proof of parity invariance}
\setcounter{equation}{0}
\hspace*{\parindent}
In this section we prove that, to all orders of the hopping parameter expansion,  the fermion propagator is parity invariant, although the action is not,  
provided that the gauge field coupling in the extra dimension is strong 
and the dynamical fermions are neglected.
\par
To begin with, let us briefly review the hopping parameter expansion. 
The lattice action
of the system with a $(2n+1)$-dimensional Dirac fermion coupled to the dynamical 
gauge field can be written as
\begin{eqnarray}
  S & = & \sum_{n,\mu} K
           [\bar{\Psi}_{n+\mu}(\lambda +\gamma^{\mu})U_{n,\mu}\Psi_{n}
     +\bar{\Psi}_{n-\mu}(\lambda - \gamma^{\mu})U_{n,\mu}^{\dagger}\Psi_{n}]\n
    &   &+\sum_{n}\bar{\Psi}_{n}\Psi_{n}\n
    &   &+\beta\sum_{(\mu, \nu)} U_{P}+\beta_{s}\sum_{(s, \mu)} U_{P}, 
\end{eqnarray}
where 
\begin{equation}
K = \frac{1}{M+\lambda}
\end{equation}
is the hopping parameter and $M$, $\lambda$ are the mass and the coefficient 
of the Wilson term, respectively. 
$\beta = 1/g^{2}$ and $\beta_{s} = 1/g_{s}^{2}$ are the gauge coupling 
constants in the $(\mu, \nu)$ plane in $2n$-dimensions and in the 
$(s, \mu)$ plane in the extra dimension, respectively.
The lattice spacing is set  unity throughout this letter. 

Let us define the gauge invariant  two point fermion Green function 
 by inserting link variables along the line 
connecting the two points as 
\begin{equation}
\langle\Psi_{n} \prod U \bar{\Psi}_{m} \rangle.
\label{green-fun}
\end{equation}
Integrating the fermion field gives a summation of 
\begin{equation}
K^{l}(\lambda \pm \gamma^{\mu_{1}})(\lambda \pm \gamma^{\mu_{2}}) \cdots 
(\lambda \pm \gamma^{\mu_{l}}) \prod U
\label{hop}
\end{equation}
over the paths connecting the two points $n$ and $m$, 
where the $\pm$ signs in front of the $\gamma^{\mu}$'s are 
determined according to the direction of the arrows along the path (Fig.1).
As for the gauge field,
 we have a product of $U$'s along the loops composed of the path 
considered in the above summation and the line originally introduce in order 
to make the fermion propagator gauge invariant. 
Integrating the link variables, then, gives a summation over the surfaces 
which have  the loop  as the boundary. 

We now prove 
the parity invariance of the Green function
\begin{equation}
\langle \Psi_{n}\prod U \bar{\Psi}_{m} \rangle =
\langle\gamma^{0} \Psi_{n'} \prod U\bar{\Psi}_{m'}\gamma^{0}\rangle,
\end{equation}
where $n'$ and $m'$ are the parity--transformed points of 
$n$ and $m$, respectively. 
This equation corresponds to eq.(\ref{eq:corrparity})
in section \ref{sec:parity}. 
We prove it by classifying the paths of the fermion line as follows.

\subsubsection*{1. The case in which the path is restricted on the boundary}

When the fermion Green function is parity-transformed, 
the corresponding fermion path in the transformed Green function 
can be obtained from the original fermion path by reversing all the 
$x^{i}$ coordinates (See Fig.2).
Since the parity transformation reverses the direction of the arrows parallel
 to the  $x^i$-axis, 
the contribution of the transformed path is given by reversing the signs in front of the
$\gamma^{i}$'s in the expression for the contribution of the original path. 
However, the reversions of the  signs in front of the $\gamma^{i}$'s 
are cancelled
when $\gamma^{0}$ in the parity transformation passes through 
the product of $(\lambda \pm \gamma^{i})$'s,
due to the anti-commutation relation of $\gamma^{0}$ and $\gamma^{i}$ as
\begin{equation}
(\lambda-\gamma^{0})(\lambda-\gamma^{i})\cdots (\lambda-\gamma^{0})=
\gamma^{0}(\lambda-\gamma^{0})(\lambda+\gamma^{i})\cdots(\lambda-\gamma^{0})
\gamma^{0}.
\end{equation}
This shows that there is a one-to-one correspondence of the paths 
contributing equally to the Green function and the parity-transformed one.
Hence, as long as we consider paths on the boundary, parity invariance 
cannot be broken.
This result can be naturally understood in the following way. 
When the fermion propagates on the boundary 
and not in the direction of the extra dimension, 
the left--right asymmetry cannot appear 
since the $2n$--dimensional action is invariant 
under the parity transformation(See section 2.).

\subsubsection*{2. The case in which the path goes around 
in the extra dimension (Fig. 3)}

Since we are working in the strong coupling case, {\it i.e.} 
$\beta_{s}=0$, the contribution of the above case vanishes after 
integrating over link variables. 
The cases to be considered in order to get a nonzero contribution are 
the followings. 

\subsubsection*{3. Tube-like propagation in the extra dimension (Fig. 4)}

We first consider the case in which the fermion propagates 
along the $s$-direction and then goes straight 
back along the same line. 
The product of $\gamma$--matrices corresponding to this tube--like propagation 
is 
\begin{equation}
(\lambda -\gamma^{2n})^{l}(\lambda +\gamma^{2n})^{l} = (\lambda^{2}-1)^{l},
\end{equation}   
where $l$ is the number of times of the hopping in the extra dimension.  
This is a c--number and the $\gamma^{0}$ in the parity transformation can 
pass freely through the above expression, which means that the 
parity invariance is satisfied for this case as well.

\subsubsection*{4. Tube--like propagation with a closed loop 
on a constant-$s$ plane (Fig. 5)}

We next consider the case in which the fermion propages along the $s$--direction, 
  moving around a closed loop on a constant--$s$ plane and 
then goes back along the same line it came along. 
Let us consider the sum of two such contributions with the following pair 
of loops, one of which is  obtained by rotating the other 
on the constant--$s$ plane around the point on the tube. 
The expression to be considered is
\begin{eqnarray}
&  &             (\lambda -\gamma^{2n})^{l}(\lambda -\gamma^{\mu_{1}})
(\lambda - \gamma^{\mu_{2}}) \cdots (\lambda -\gamma^{\mu_{m}})
(\lambda + \gamma ^{2n})^{l}\n
&+ &             (\lambda -\gamma^{2n})^{l}(\lambda +\gamma^{\mu_{1}})
(\lambda + \gamma^{\mu_{2}}) \cdots (\lambda +\gamma^{\mu_{m}})
(\lambda +\gamma ^{2n})^{l},
\end{eqnarray}
where $(\lambda \pm \gamma^{2n})^{l}$ 's come from the tube--like 
propagation along the $s$-direction and $(\lambda \pm \gamma^{\mu})$ 's between them
 come from the  propagation along the  loops on the constant--$s$ plane. 
Note that the signs in front of $\gamma^{\mu}$'s  are 
opposite due to the definition of the pair of loops. 

We expand the product of the $(\lambda \pm \gamma^{\mu})$'s into the sum of 
$\lambda^{m_{1}}\gamma^{\nu_{1}} \gamma^{\nu_{2}} \cdots \gamma^{\nu_{m_{2}}}$,
where $m_{1} + m_{2} = m$.  
When we consider a term where $m_{2}$ is odd, the two contributions 
cancels. 
When $m_{2}$ is even, the tube part $(\lambda \pm \gamma^{2n})^{l}$ 
can pass through this loop part 
$\gamma^{\nu_{1}} \gamma^{\nu_{2}} \cdots \gamma^{\nu_{m{2}}}$, 
and these tube parts become a c-number, commutable with $\gamma^{0}$.
Hence, these paths contribute equally to  the Green function
and the parity-transformed one as in the case 1, and parity invariance is 
satisfied as well.

\subsubsection*{5. Many tubes and loops (Fig. 6)}

By induction, we can show parity invariance is satisfied as well.
%
This completes the proof of the parity invariance of the fermion 
Green function. 

\vspace{1cm}

\section{Summary and Discussion}
\setcounter{equation}{0}
\hspace*{\parindent}
In this letter we have shown that the dynamics of the gauge field 
renders the fermion propagator vector--like in the strong 
 gauge coupling limit in the extra dimension provided that  
dynamical fermions are neglected. 
The result is consistent with the mean field analysis.

When we include the effect of dynamical fermions, we should take account 
of such diagrams as Fig. 7, which may make the fermion propagator 
not exactly parity invariant. 
However, since adding dynamical fermions amounts to adding various types 
of Wilson loops to the gauge field action, 
we may say that our argument   holds for such models that do not 
have Wilson loops including links in the $s$--direction after 
integrating dynamical fermions. 
We should also remind the readers that the parity invariance of the fermion 
propagator is much stronger than the fermion being vector--like. 
We might, therefore, expect that the disappearance of 
the chiral zeromode 
occurs generally in the strong coupling region of the gauge coupling 
in the extra dimension. 
Although our argument confirms that the strong coupling scenario of 
constructing chiral gauge theory seems unlikely to work, 
it does not exclude the other possibility in the weak coupling region 
explained in the Introduction. 
\vspace{1cm} 

We would like to thank Prof. S. Aoki and Dr. Y. Kikukawa 
for stimulating discussion and Prof. H. Kawai for useful comments. 
%

\end{document}